# What Isn't Complexity?

Christopher R. Stephens

$C_3$ Centro de Ciencias de la Complejidad and
Instituto de Ciencias Nucleares,
Universidad Nacional Autonoma de Mexico
Circuito Exterior, A. Postal 70-543

## 1. Introduction

The question "What is Complexity?" has occupied a great deal of time and paper over the last 20 or so years. There are a myriad different perspectives and "definitions" [1, 2, 3] but still no consensus. But what does the question mean? What do we expect from addressing it? Many think of the goal as finding an intentional definition, whereby necessary and sufficient conditions are specified, according to which anything can be uniquely classified as complex or not. On the other hand, an extensional definition takes a more phenomenological approach, characterizing the set of complex systems by trying to name its members. The intentional route faces the difficulty of either being too restrictive or too general. For example, the notion of computational complexity [4] is mathematically quite rigorous but is too restrictive and, given that maximally complex things are random bit strings, certainly does not capture the intuitive notion of what complexity is. On the other hand, defining complex systems as having many degrees of freedom and non-linear interactions is completely vacuous given that, basically, everything is like that, from a salt crystal to a zebra or from a simple atom to the human brain. One cannot argue that these conditions are not necessary, but they are certainly not sufficient. However, they do indicate two features that we should be aware of – What are things made of? and, What are their interactions?

The extensional definition runs into the problem of having to decide which systems are in the set of complex systems and which not. Unfortunately, there is a definite, subjective disciplinary bias in answering this question, a physicist and a biologist often having quite different perspectives. However, instead of trying to determine a precise boundary to the set of complex systems we can take a more pragmatic approach of starting off by listing some systems that "everyone" would agree are complex, and therefore in the set of complex systems, and others that "everyone" would agree are not complex, and therefore not in the set. We can then try to determine what properties discriminate between what we have definitely put in the set and what we have definitely left out. In taking this approach it will be useful to also pose the complementary, question: What isn't Complexity? For instance, complexity isn't many degrees of freedom and non-linear interactions. If it were, then basically everything would be a complex system and it wouldn't make any sense in trying to distinguish complexity as something different.

## 2. An example of a Complex System – Human language

As stated, instead of first making an a priori definition of complexity and complex system, we will take the path of first considering some systems that definitely would be accepted (at least by the vast majority) as being complex and compare them to some examples of systems that definitely would be accepted (at least by the vast majority) as being not complex. So, starting off with what is complex – two classes of system that immediately spring to mind, that certainly intuitively qualify as complex, are biological systems and human languages. We can think of the former as representing "physical" complexity and the latter "symbolic" complexity. Let us address first the question of human language.

*To be, or not to be--that is the question:*
*Whether 'tis nobler in the mind to suffer*
*The slings and arrows of outrageous fortune*
*Or to take arms against a sea of troubles*
*And by opposing end them. To die, to sleep--*
*No more--and by a sleep to say we end*
*The heartache, and the thousand natural shocks*
*That flesh is heir to. 'Tis a consummation*
*Devoutly to be wished. To die, to sleep--*
*To sleep--perchance to dream: ay, there's the rub,*
*For in that sleep of death what dreams may come*
*When we have shuffled off this mortal coil,*
*Must give us pause.*

Figure 1: Hamlet's famous soliloquy from Shakespeare's play of that name

In Figure 1 we see an excerpt from Hamlet's famous soliloquy in the Shakespeare play of the same name, while in Figures 2 and 3 we see two sequences extracted from a protein and the human genome respectively. There are several questions we can pose: are all three symbolic sequences complex? If so, is one more complex than the other? Naturally, our understanding of the text is greater - we speak the "lingo" – meaning that we understand both its grammatical and semantic content.

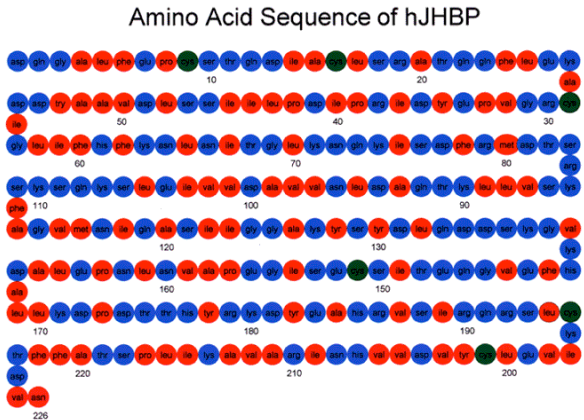

Figure 2: An amino acid sequence from a protein

Figure 3: A nucleotide sequence from the human genome

Imagine for a moment though, that Figure 1 is a mystery, as to a large extent are the others. We will analyze it from the point of view of a statistical physicist. What would probably strike one first, at least in the case of non-pictographic languages like English, is that there is a fundamental "microscopic" degree of freedom – the "letter" (by which we also include some other elementary symbols such as a space, number, punctuation marks etc.) and that language consists of an ordered sequence of such letters. The first question one might ask concerns the relative frequency of the different letters, $<x_i>$, in a given sample of text, noting that these are highly non-random. For instance, in English, the vowels "a" and "e", with frequencies of about 8% and 13% respectively, are much more common than consonants, such as "x" and "z", which have frequencies of 0.15% and 0.07%.

What would we do next – look for correlations? Taking text as a linear ordered sequence, we can look for correlations between pairs of letters located at positions l and l' in the sequence and evaluate the two-point correlation function $<x_i(l)x_j(l')>$, which will be a function of (l-l') only. What would we observe? In the case of l' = l+1 we are considering digraphs; then, we would find, for instance, once again in English, that $<t(l)h(l+1)>$ >> $<q(l)z(l+1)>$ due to the fact that "h" often follows "t" but "z" does not follow "q", "th" being the most common digraph in English. We'd also find that the correlations are non-commutative, in that $<x_i(l)x_j(l')> \neq <x_j(l)x_i(l')>$, e.g., "th" is much more common than "ht". Proceeding, we could consider three-point functions, $<x_i(l)x_j(l')x_k(l'')>$, finding for l' = l +1 and l'' = l + 2 that the trigraph "the" was the most frequent. What is more, we could observe that frequently the trigraph "the" has a space symbol on either side in the linear sequence, as had other combinations of letters.

A perspicacious statistical physicist, with an insight into how effective degrees of freedom that describe collective behavior can emerge from combinations of underlying microscopic degrees of freedom, might be led to posit the existence of an effective degree of freedom of the type, "SPACE $x_i(l)x_j(l+1)…x_k(l+n)$ SPACE", and might be led to call this new emergent degree of freedom a "word". The most common word in English is "the" which comprises about 2% of all words in a typical text. Armed with this knowledge of the existence of a "bound state" of letters – the word – our inquisitive statistical physicist following the lead of an analysis of the relative frequency of different letters might be inclined to now think about the relative frequency of these words. In Figure 4 we see a plot in log-log coordinates of the frequency of a word versus its rank for Wikipedia web pages. Lo and behold, one finds a straight line signifying that the relation between word frequency as a function of frequency rank is a power law of the form $f(r) = A/r^a$, where a=1 for $r < r_c$, and $r_c \sim 10,000$. This is the famous Zipf's law [5]. Interestingly, there appears to be a crossover at $r_c$ such that for $r > r_c$ $f(r) = B/r^2$. This is all very exciting and surprising for our statistical physicist. The question is: does it tell us anything about language and its associated complexity?

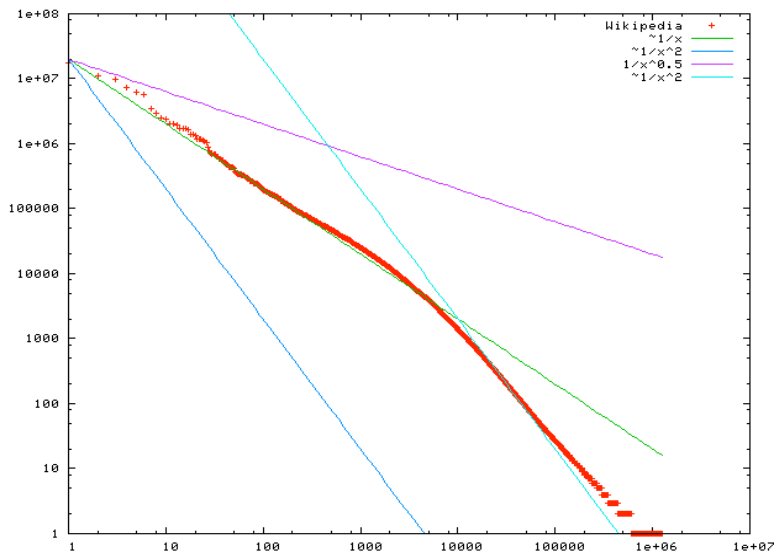

Figure 4: A plot of word frequency in Wikipedia (November 27, 2006). The plot is in log-log coordinates. *x* is rank of a word in the frequency table; *y* is the total number of the word's occurences. Most popular words are "the", "of" and "and", as expected. Zipf's law corresponds to the upper linear portion of the curve, roughly following the green *(1/x)* line (Victor Grishchenko)

Now, all this analysis for text in natural language can also be carried out for the systems of Figures 2 and 3. In fact, many of the results are somewhat analogous. Genes, indeed, are somewhat analogous to words. Beyond that though, we really don't have much understanding of the effective degrees of freedom of genetic systems. In fact, there is currently ample debate as to what degree a gene is an isolatable effective degree of freedom [6].

## 3. Bored on the "Edge of Chaos"

Having seen the appearance of a power law and an associated scale invariance in the context of human languages our statistical physicist is reminded of what has been an important concept in the study of complexity - the "Edge of Chaos" [7, 8]. The idea is that complex systems, in analogy to the porridge of the three bears, are neither too "ordered" nor too "disordered" but finely balanced between the two. The original application of the "Edge of Chaos" idea was to the very interesting subject of information processing in cellular automata, where it was shown that, in order to promote "complex" computation, an adequate balance between information storage and information transmission was required; information transmission requiring a low entropy (low noise) regime while information storage, which required many available states, is associated with a high entropy regime.

Another very interesting example of a balance between ordering and disordering tendencies is associated with the "error threshold" [9] in molecular evolution, where there can be a fine balance between mutation and selection. If the mutation rate is too high then selection cannot act, while if it is too low the system cannot easily evolve. There is, in fact, evidence that some viruses do operate with mutation rates near the error threshold [10] in order to strike a balance between exploration and exploitation. In these different examples, as in the case of word frequency and Zipf's law, one encounters power laws characterizing the relationship between certain variables. These power laws are a consequence of the absence of a characteristic scale and the associated phenomenon of scale invariance.

However, this idea of balancing order and disorder is an old one in statistical mechanics, being associated with the idea of a phase transition, especially a second order one, where physical quantities exhibit power law behavior, just like Zipf's law, and the system displays "criticality". For instance, the magnetization of a ferromagnet, $\varphi$, is related to the temperature deviation from the critical temperature, $t = T_c - T$, as $\varphi \propto t^\beta$, where $\beta$ is a characteristic critical exponent. Near the critical temperature there is a fine balance between the disordering tendencies of thermal fluctuations and the ordering one due to the fact that the magnetic dipoles in the ferromagnet energetically prefer to be in an aligned state. In this case, the temperature has to be tuned quite delicately to be in the vicinity of the critical temperature in order to see scale-invariant power law-type behavior. This requirement of tuning is very common.

There are systems, though, where the analog of the temperature tunes itself without external influence. These systems exhibit self-organized criticality, and the most famous example is that of a sand-pile [11]. One imagines slowly dropping sand grains onto a pile on a table. The pile eventually reaches the edge and a dynamic equilibrium is reached whereby the amount of sand arriving and the amount leaving are equal. In such circumstances the angle between the sides of the pile and the table is fixed. On the sides of the pile there take place avalanches of sand grains. The frequency, f, of such avalanches depends on their size, n, such that $f(n) \propto n^{-\alpha}$. Once again, we see the ubiquitous power law behavior characteristic of scale invariance. Similar results exist for earthquakes, species extinctions, fundamental laws of physics, such as electromagnetism and gravity, metabolic rates, city populations, income distribution, flying velocities of insects and stock prices, to name just a few!

What are we to make of a phenomena that is so omni-present? Is it something very deep or very trivial? Of course, no one could deny that on a case by case basis all these phenomena are extremely interesting, but remember, we are here trying to find discriminating characteristics of complex systems. The ubiquity of power law behavior, applying equally to "simple" systems, such as ferromagnets and Newton's universal law of gravitation, as to "complex" ones, such as cities' populations and stock prices, puts the "Edge of Chaos" in its most general meaning in the same class as other characteristics, such as "many degrees of freedom" and "non-linear interactions", vacuous as a means of saying what truly discriminates complex systems from simple ones. Thus, although the "Edge of Chaos" may be a characteristic of complex systems, it is not a defining characteristic, in that it does not discriminate between what is definitely complex and what is definitely not complex.

I mentioned above that there are certainly disciplinary biases associated with considerations of what is complex and what not. This is also true for whether or not one thinks that scale invariant behavior is normal, or surprising, or not. For particle physicists, what is surprising is the existence of masses, i.e. preferred scales, rather than scale invariance. For a solid state physicist however, seeing scale invariant behavior near a second order phase transition seems more surprising in a system that possesses important fixed scales, such as the inter-atomic spacing in a crystal lattice.

## 4. Structural Building Block Hierarchies

So, not everything exhibits power law behavior. On the contrary, there are just as many instances of effective degrees of freedom that do have a characteristic scale – massive elementary particles, nucleons, nuclei, atoms, molecules and macromolecules, cell nuclei and other organelles, cells, tissues and organisms. For instance, there is no scale invariant distribution of elephant sizes, or number of nucleotides in the DNA of a given species. In particular, systems that manifestly exhibit physical complexity, such as biological organisms, are associated with a myriad of quite distinct characteristic scales, at each of which the characteristic effective degrees of freedom are radically different. What is more, such effective degrees of freedom exhibit a hierarchical relationship – loosely speaking, nucleons are composed of quarks, nuclei of nucleons, atoms of nuclei and electrons, molecules of atoms, cell nuclei and other organelles of biological macromolecules, cells of organelles, tissues of cells and organisms of tissues.

There is a very powerful, simple argument as to why nature should favor such hierarchical structures. If we try to construct a human by throwing atoms together we would be waiting a very long time. However, if we try to form molecules by throwing atoms together then that is much more feasible. Then, we can try and form polymeric macromolecules from encounters between their monomeric constituents, and so on. In other words, it is much easier to construct a more complex system via interactions between a not too large a number of slightly less complex potential "building blocks", than from a very large number of very simple "building blocks". One then develops a nested hierarchy of building blocks at different scales such that blocks at one scale are constituents for blocks at a larger scale, which in their turn etc. Nucleosynthesis in the sun works in such a way. A salient feature of building blocks is that they have characteristic sizes. I know of no manifestly complex system that is constructed from scale invariant blocks.

Basically all progress in science has come from being able to study systems by restricting attention to phenomena associated with a fixed scale, thereby isolating a given characteristic type of building block whose interactions can be studied using an appropriate effective theory at that scale. In fact, the compartmentalization of science into a large number of separate sub-disciplines, such as particle, nuclear, atomic and molecular physics, are a direct consequence of this property of nature. The question is: to what extent can complexity be understood by taking this approach?

To think on this further, let's return to human languages. Certainly, languages are associated with a hierarchy of building blocks. The most basic, in written language, is the letter, followed by syllable, then word, then phrase, then sentence, then paragraph etc. These different building blocks are certainly associated with characteristic sizes. The sizes of words and sentences certainly do not follow power law distributions. A characteristic of language then is the existence of non-trivial structure on many different scales, each scale being associated with a different effective degree of freedom or building block. Can we now argue that the presence of a hierarchy of building blocks with characteristic scales is a sufficient condition for complexity? Certainly, this seems to be a much more discriminating criterion than "many degrees of freedom", or "non-linear interactions", or "Edge of Chaos". But is it enough? Certainly biological systems and human languages exhibit such hierarchies. What about other systems?

Most people would not consider a hydrogen atom to be a complex system. A hydrogen atom is a bound state of a proton and an electron. The proton is, in its turn, a bound state of quarks. Perhaps, these in their turn are low energy excitations of a more fundamental degree of freedom, such as a string or a membrane. So, even at the level of the lowly hydrogen atom, there exists a hierarchy of building blocks. Are building block hierarchies a red herring then? In the introduction, I mentioned that, although many degrees of freedom and non-linear interactions in no way indicated complexity, they were associated with two important properties – what elements are things made of and how do those elements interact. We are trying to posit that complex systems are composed of hierarchical building blocks, but have hit the obstacle that some non-complex systems also exhibit such hierarchies. So, we pose the question: Do these hierarchies exhibit similar interactions in complex and non-complex systems?

## 5. Hierarchical interactions

If we consider the air that you are currently breathing, it exhibits a hierarchy of effective degrees of freedom. It is composed of a set of mainly simple molecules, principally nitrogen and oxygen. These in their turn are composed of pairs of covalently bonded oxygen and nitrogen atoms, which in their turn are composed of a fixed number of electrons and nucleons of two types – protons and neutrons. The nucleons themselves are composed of a fixed number of quarks. Thus, in this system there is a hierarchy of effective degrees of freedom, or building blocks, passing from one scale to another. The air as a non-ideal gas in approximate thermal equilibrium represents the ultimate level of aggregation for this system.

In terms of interactions, the energy scales that measure the degree of interaction between different building blocks are much higher at the level of quarks than they are at the level of nucleons, which in turn are higher than the electromagnetic interactions between nucleus and electrons, which in their turn are higher than the covalent interaction between the atoms that make a molecule, while the Van der Waals interactions between the molecules themselves are even weaker still. Thus, the interaction strength systematically weakens as we go to larger scales. Put another way, it is easier to separate two nitrogen molecules, than two oxygen atoms, while separating the electrons is even harder and the nucleons even more so. Interestingly, the interaction strength at the level of quarks in a nucleon is so high that they cannot be separated at all without creating new ones! We can also think of the degree of interaction in terms of the ease with which the system can be perturbed. Separating the constituents is obviously one way to do this.

As another example, consider a spin chain, where, for simplicity, we take the spins to be binary valued, as in the Ising model, though any alphabet would do - 4 for the nucleotides of DNA, or 32, thinking of the English alphabet along with a space symbol and five punctuation marks. We consider nearest-neighbor ferromagnetic interactions so that aligned spins is the energetically preferred state. In this system there are two ways of characterizing the degrees of freedom – in terms of the individual spins, or in terms of domains as contiguous sets of aligned spins. Obviously, a domain can be simply represented in terms of the underlying spins. The same is true of a sand pile, where the analogs of spin and domain are sand grain and avalanche. In this case there is no hierarchy or, rather, the hierarchy is of "depth" two in scale – the domain/avalanche effective degree of freedom being composed of the microscopic spin/sand grain degrees of freedom. The domains/avalanches are building blocks for the whole system – sand pile/spin chain - but in between the micro and the macro there are no more intermediate layers. Of course, if we went beyond the above description, and considered how the above abstract spins actually represent properties of atoms and that, again, atoms are composed of electrons and nucleons etc. then we would see a richer hierarchy.

So, how do these hierarchical interactions differ from those of a complex system? Language clearly is also composed of a rich hierarchy of different building blocks. How do these blocks interact? What will be our notion of interaction strength? We can consider different complementary measures. We could for instance, just demand that the words in the system exist in a given lexicon irrespective of grammar or semantics, for example giving a fitness f to a word that exists and a fitness f' << f to one that doesn't. We could demand that the words obey the rules of grammar without necessarily being in the lexicon, such as occurs in nonsense verse. We could also demand that the words are fully consistent both grammatically and semantically, this corresponding to the lowest energy or highest fitness state.

Let's consider a famous sentence in English, from the point of view of a statistical physicist, not a linguist, I hasten to add. "*The quick brown fox jumped over the lazy dog.*" contains all the letters of the English, alphabet so we can see all the microscopic degrees of freedom displayed. This sentence is completely self-consistent, both from the grammatical and semantic point of view.

We can note that there are certainly strong "interactions" between letters within words. "The quick brown ofx jumped over the lazy dog" doesn't makes sense as "ofx" does not exist as a word in the standard English lexicon. However, it does make good grammatical sense if we accept "ofx" as a noun. Words are, in general, very sensitive to the introduction of noise by "mutating" letters. A mutation of "fox" to "fzx" leads to something not in the lexicon. However, a mutation to "fix" gives something that is in the lexicon. Indeed, different words have different degrees of robustness in the presence of mutations. For instance, the word "dead" is much more robust to mutations – bead, deed, lead etc. - than the analogous word "defunct". It is really the constraint that a word is in the lexicon that makes the interactions between letters in a word so strong. The corresponding fitness landscape for letter combination is rugged but not random as there are preferred non-random combinations of letters that accord with the sounds that the human voice is capable of making.

Passing now to words: any permutation of the sentence is fine if we are only concerned with having words in the lexicon. In this case there are no interactions between the words. However, the constraints of grammar induce interactions between them. Thus, if we think of grammatical correctness as a measure of fitness then: "The quick, brown jumped fox over the lazy dog." is of lower fitness than the original sentence as the placement of the verb "jumped" is grammatically incorrect, thereby indicating that the fitness of the sentence is sensitive to the order of the words. On the other hand, the permutation: "The lazy dog jumped over the quick, brown fox." is perfectly fine grammatically, though it does sound a little counterintuitive, while "The quick, brown lazy jumped over the dog fox." makes no sense as we have only a string of adjectives in the place of the subject of the sentence.

There are two important points that emerge here: first, that there are different *classes* of words – nouns, verbs, adjectives, prepositions, articles etc. and second, that word position matters. The degree of interaction between two words in a sentence depends sensitively on both their type and position. Once again though, we must ask: Is this different to what we see in non-complex physical systems? After all, type and position often count for a lot in physics too. Back to our considerations of air! Interchanging the positions of two nitrogen molecules in the air makes no difference to its macro-state. Nor does interchanging the positions of an oxygen and a nitrogen molecule. However, if we interchange a neutron in one of the nitrogen nuclei with a proton in another, we now get a carbon 14 nucleus. This would make an important change in the energy of the nucleus which would affect its interactions at higher levels, though clearly it wouldn't affect the macro-energy state. So, what are we missing? Or, maybe, language isn't complex after all.

## 6. Hierarchical Emergence

Above, I argued that lexicographic and grammatical constraints could be viewed as imposing interactions between letters and words respectively. For words, grammatical considerations essentially restrict the interactions to phrases and sentences. Are there "longer range" interactions? Yes. And these interactions are associated with an important aspect of language that we have not much touched on up to now – semantics. We have assumed that our erstwhile statistical physicist is ignorant of any meaning associated with any particular symbol strings, even though he/she has been able to determine an extremely non-trivial structure with the emergence of many different collective degrees of freedom or building blocks. Semantics is to do with the map between a particular ordered sequence of symbols and an associated set of concepts. "The quick brown fox jumped over the lazy dog." has a clear semantic meaning for any English speaker. This is not so for a non-English speaker. Thinking of both as statistical physicists however, the non-semantic statistical content is identical. So, can the semantic content be thought of in statistical terms? Yes. The semantic content is associated with a different set of correlations than those intrinsic to the letter sequences themselves – correlations between letter sequences and concepts. The sequence "fox" refers to an animal with a certain set of phenotypic characteristics. I would bet that the patterns of neural activity are quite distinct between an English and non-English speaker when presented with this sentence.

Semantics induces interactions beyond the scale of a sentence. Consider:

1. **"The quick, brown fox jumped over the lazy dog. The dog woke up, startled."**

2. **"The quick brown fox jumped over the lazy dog. Please pick up milk on the way home from the office."**

In the first combination, both sentences are grammatically and semantically self-consistent in themselves. However, the second sentence is also a continuation in the narrative associated with the first. For that reason there is an interaction between the two. For the second combination, once again, both sentences are grammatically and semantically fine. In this case though, the second sentence is completely logically separate from the first in a semantic sense and therefore there is no interaction between them.

But now, having arrived at the scale of two sentences we begin to come to the crux of the matter, where we can see some property that human language, as a representative of the set of complex systems, has that other simpler systems don't. The semantic content of the two sentences is a multi-scale phenomenon. What do I mean by that? Each word has a semantic content, as does each phrase and each sentence (and each paragraph, section etc.). If we think that every one of these building blocks has its own characteristic scale, then the full semantic meaning involves an *integration* across all these different scales. If we restrict our attention to the building blocks of any one scale without knowing how those blocks integrate into others at a higher scale then we lose the full semantic meaning.

For instance, just looking at combination 1) above; at the level of individual words there are two nouns, 3 adjectives, one verb, one preposition and two articles. We can see that we are talking about, a fox ("Any of various carnivorous mammals of the genus *Vulpes* and related genera, related to the dogs and wolves and characteristically having upright ears, a pointed snout, and a long bushy tail."), a dog ("A domesticated carnivorous mammal *(Canis familiaris)* related to the foxes and wolves and raised in a wide variety of breeds."), the action of jumping ("To spring off the ground or other base by a muscular effort of the legs and feet."), something that is lazy ("Resistant to work or exertion; disposed to idleness.") etc. Looking at just the words we have no means to extend the semantic content to higher order building blocks. In terms of phrases, 1) has four: "The quick, brown fox" is a noun phrase, "jumped" a verb phrase, "over" a prepositional phrase and "the lazy dog" another noun phrase. Irrespective of how these phrases are joined together as building blocks to form a sentence we can see that there is now new semantic content at this scale. The first noun phrase tells us that it's the fox that is quick and brown while the last one tells us that it is the dog that is lazy. Apart from that we still don't know who jumped over whom. If we go up to the sentence level of building block however now the ordering of the phrases tell us that it was the quick brown fox that jumped over the lazy dog. A similar analysis of the second sentence tells us that it was the dog that woke up and that its condition was one of "startled". However, we can now do one further integration step and combine the two sentences together to understand that the dog's waking up and startled state was a direct causal result of the fox jumping over it.

To recap: what distinguishes language as a complex system isn't just that there is a hierarchy of building blocks but, rather, the existence of emergent properties that depend on all levels of this hierarchy. Meaning is such a property that transcends any given fixed level of building block. It is present at every level to a given degree but at each higher level a new integrated form emerges that depends on the lower levels. Thus, the meaning of a word depends on the letters that form it; the meaning of a phrase depends on the meaning of the words that it is composed of; the meaning of a sentence depends on the meaning of the phrases that constitute it. There is no meaning to a building block at a given level without having the meaning of its constituent lower level building blocks. It is this that I claim is a true hallmark of complexity.

What about physical complexity? Biological systems clearly display hierarchical building blocks. Are they more like language or more like air? The case of language illustrated that it was not the building block hierarchy per se that was associated with complexity but rather how a particular "observable" – meaning – induced interactions between different levels of that hierarchy. I would argue that there exists at least one "observable" in biological systems that has the same property – "fitness", thought of as reproductive success. With meaning I argued that it was manifest at every building block level, but became transformed passing from one level to a higher one. Clearly, with meaning the whole is not the sum of the parts, even though the parts clearly contribute. Fitness in the same way has a meaning at each level but, again, the whole is not just the sum of the parts.

To give an example: start with a microscopic scale, that of the important biological macromolecules, such as DNA, RNA, proteins etc. These obviously contribute to fitness. There are many micro-events, such as mutations, that have important macro-consequences, such as sickle-cell disease. There, a mutation alters the nature of haemoglobin proteins so that red blood cells form with anomalous sickle-like shapes which change their oxygen carrying capacity. However, the consequences are restricted in their scope. The sickle-cell mutation does not directly affect other body functions but does affect overall fitness as it reduces life expectancy in non-malarial environments. Essentially, cells with a gene with the sickle-cell mutation aren't doing what they should and this can be reflected in a lower contribution to the overall fitness of the organism from the cell reproduction mechanism associated with those cells. This in turn can be reflected at a higher level by a lower fitness contribution from the red blood cells themselves, which, at an even higher level, can be thought of as a lowered contribution from the red bone marrow that produces them and even up to a lower contribution from the entire cardio-vascular system.

Contributions to fitness originate in all building block scales, just as contributions to meaning do in the case of language. In fact, it is probably not too fanciful to think that meaning in language is closely associated with an analogous concept of fitness or utility, as would accrue, for example, from being warned by a companion of a risk, such as a dangerous predator. In more general terms utility can stem from many things: following a cooking recipe, building an engine or solving differential equations, even being entertained by a book.

# 7. The tyranny of physical law and the difference between "being" and "doing"

Besides being characterized by a rich hierarchy of building blocks at different scales that interact in an integrated fashion, what else characterizes complex systems? Once again we return to the logic of what properties do manifestly complex systems exhibit that one does not tend to find elsewhere? In physics and chemistry we have worked within the same paradigm for nearly 500 years – that the world can be described by a state and that there is a unique dynamical law that evolves that state in time. What has changed over time is our understanding of that unique dynamical law, passing from the laws of classical mechanics to those of quantum mechanics and incorporating relativity. In fact, not withstanding the quasi-theological pursuit of a "theory of everything" one could sensibly take quantum electrodynamics as the fundamental microscopic theory from which just about everything up to terrestrial scales should be derivable. Biological systems in principle fall into this category. Of course, no one seriously believes that quantum electrodynamics can be sensibly used to understand biology. It is not facile, however, to ask why not? Am I hidden as a collective excitation somewhere deep in the generating function for quantum electrodynamics? Given that every atom that composes an amoeba has to obey quantum electrodynamics then it follows that the amoeba itself obeys quantum electrodynamics. The latter is sufficient to describe and predict the behaviour of individual atoms, but not the enormously complex dynamics of the huge number of interacting macromolecules that characterize the amoeba.

But is it just a question of complication? For example, the energy levels of hydrogen can be solved for analytically – at least in the confines of an approximate model – the one particle Schroedinger equation. The energy levels of Uranium on the other hand cannot. However, numerically, the levels of Uranium can be solved for very accurately. So, what does an amoeba do that a uranium atom doesn't? The question itself, in fact, contains the answer – the amoeba "does" things. An atom passively obeys Schroedinger's equation, while a tennis ball passively obeys Newton's laws. A cat necessarily obeys both - but not passively. A cat falling upside down will right itself in mid-air so as to land on its feet. Both cats that fall on their head and cats that land on their feet obey all the restrictions imposed by the laws of physics and chemistry. Clearly then, these cannot help us decide why we only see cats that land on their feet. They cannot be used to distinguish between head-first and feet-first cats. The point is, the cat or the amoeba have a huge number of potential internal states that are all equally compatible with the fundamental laws of physics. What is more, that internal state can change dynamically in reaction to what is happening in the environment. A word that can be applied to such dynamically changing internal states is "strategy".

Two competing strategies that falling cats can adopt are: "head-first" strategy and "feet-first" strategy. As the former leads to more injuries and a lower survival rate, the latter strategy has been selected as a useful adaptation. What is a strategy in this context? It is a rule for updating the state of the system. The fact that different strategies exist means that there are different dynamical rules for updating a state. This also seems to be an important characteristic of complexity in the biological regime: That a more adequate description of the system's dynamics is in terms of both states and update rules or strategies.

Is this phenomenon only present in biological systems? A wonderful illustration of it in an artificial life, as opposed to real life, context, can be found in Karl Sims' [12] stunning work on evolving virtual "creatures" to do different tasks. What Sims did was to create a creature by combining a simple morphology, obtained by combining together a hierarchy of three-dimensional rigid parts – quite literally building "blocks". These building blocks were represented "genetically" as a directed graph, where each node in the graph contains information about the dimensions of the corresponding block and how it is joined to its parent block. A neural network was used as a brain that indicated how one block should move relative to another. The creatures were set a task – swim in a virtual sea, or move on a virtual surface etc. A population of such creatures was then evolved over time to see what strategies emerged for tackling the given problem. In Figure 5 we see a one such evolved solution for swimming. Go to http://www.archive.org/details/sims_evolved_virtual_creatures_1994 to see the impressive movie.

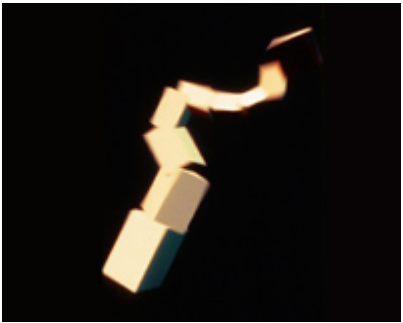

Figure 5: A virtual creature evolved for swimming in a virtual sea.

This represents much more than just a striking visualization. Great care and attention went into this work to assure that all the elements of the simulation were physically sound, i.e. that both internal movement and movement through the media were all completely consistent with the dynamics of rigid bodies in viscous media or in a gravitational field. Why do I take pains to emphasize this? Because the results of this simulation - the evolution of fitter and fitter organisms adapted to a given environment – *cannot be understood* in terms of the laws of physics

What do I mean by this? The space of possible strategies for a given complex system is, of course, restricted by the laws of physics. For instance, humans cannot fly unaided because, among other things, we cannot generate enough lift without wings. But the laws of physics only bound what the possibilities are, not which ones are chosen or why. This is completely different to what happens for physical systems themselves. In that case the tyranny of the laws of physics is such that there is only one relevant update rule, either at the micro- or the macro-level. Thus, a molecule, a pollen grain or a baseball are all constrained to obey update rules that tells us what their state will be later given that we know it now. It is for that reason there is no meaning to the concept of adaptation in the physical world. Adaptation is to do with being able to adopt a new update rule (strategy) within a given environment.

Physical systems only have to do one thing – "be". There is no requirement that an atom evolve in order to survive in a hostile environment. When air is heated to sufficient temperature, the atoms that form the molecules in it will start to dissociate. The molecules do not develop an adaptation enabling them to survive in this hostile thermal environment. If we keep raising the temperature we will start to ionize the atoms, separating off some of the electrons. There is no possibility of an adaptive change in the atoms that enables them to stop losing electrons. However, if it gets cold I put on a coat. The origin of this is that in human prehistory, those who could prevent heat loss by covering themselves would have higher survival rates than those who didn't. Just as with the falling cats, there are two strategies – cover up and reduce heat loss, or stay uncovered and risk hypothermia. Unlike the atoms, a change in environment in this case can lead to more than one response.

The emergence of strategy as a description of a complex adaptive system is a truly emergent phenomenon. But, where does it emerge? If we think of chemical evolution, say in the context of an RNA world, where, as an example, different RNA molecules can "compete" in the context of the enzymatic catalysis of a particular interaction, it does not seem very natural to think of the RNA molecules as having a "strategy". At this level, the tyranny of the laws of physics is still manifest.

As emphasized physical systems only have to do one thing – "be". They do not have "choices". Biological systems however do have choices, different strategies, all consistent with the restrictions imposed by the laws of physics. What is more, different strategies are often associated with different elements and levels of the building block hierarchy, all of them exquisitely choreographed by the requirements of evolution to work in harmony.

This discussion naturally leads us to consider - what is adaptation? Does a population of RNA molecules "adapt" searching for an optimal configuration for catalyzing some reaction? We can paint this at another level, in the context of artificial evolution, such as in Genetic Algorithms or Artificial Life. In Genetic Algorithms, the relevant dynamical equations are almost identical to those long familiar from population biology, where a population of chromosomes evolve under the action of selection and genetic mixing operators, such as mutation and recombination. The equations that direct the dynamics in these systems is very familiar from physics and chemistry: Stochastic dynamics of Markov chains in the case of finite populations, while in the infinite population limit, the relevant equations are a set of deterministic, non-linear first order difference equations – fiendishly difficult to work with, but conceptually within the same paradigm as any other traditional "differential equation" based approach. But it is precisely such systems that I argue are subject to the "tyranny of law". In the case of a Genetic Algorithm, the restrictions do not come from any physical requirement but rather the restrictions of the model itself.

Is a Genetic Algorithm really adapting or, more provocatively, is population genetics a model for adaptation? In both cases the dynamics takes place in a space of states, but there is no explicit dynamics for any update rule. Rather, an explicit "fitness" function (viability in the context of population genetics) is used as a proxy for how well the system does. In Genetic Algorithms this paradigm has been successfully used in the context of combinatorial optimization, where there does exist a concept of "best" solution and it is reasonable what a solution is rather than what it does. In the more interesting case of Genetic Programming, this is not the case, an evolved solution representing a computer program and therefore a solution that does something as opposed to just is. For these reasons I would argue that biological adaptation really should be thought of as taking place in the space of strategies and states, not in the space of states alone. In fact, I believe that adaptation cannot emerge from any paradigm where a fitness function has been specified a priori.

# 8. "Specialist" vs. "Generalist": Evolution as the development of multi-tasking

A key element of biological systems then is that their dynamics is naturally described in a space of states *and* update rules. This, of course, is reminiscent of game theory [13]. In the latter, payoffs are assigned to individuals or groups implementing a certain strategy in the context of other players playing their strategy. A familiar example would be the children's game "rock-paper-scissors". For two players, a payoff of 1 would be given to a player playing rock, while 0 would be given to the player playing scissors. It is not clear how useful the game theory paradigm is for physical complex systems. First of all, there is the question of the tremendous number of possible strategies. Worse still, we have no idea in what space we're working in. How can we talk about the utility of one strategy versus another when we have no idea about the possibilities? Imagine 3.5 billion years ago trying to wonder about how to assign a payoff to "lion", "cockroach" or "oak tree" as the analogs of rock/paper/scissors! How would one begin to even imagine the possibility that such strategies could exist? The only option that we have is to observe the discovery of certain regions of the space of possible strategies as evolution moves forward and infer their relative benefits from observation.

We also have to address the question of strategy to do what? Game theory is usually couched in the language of one strategic objective, where there is a clear notion of winning or losing. One could argue that survival in an evolutionary context is a game. However, there is a very important difference between evolutionary survival and "rock-paper-scissors" that can be couched in the following way: Really, an organism is involved in a vast number of games and has to develop strategies for all of them. It also has the pressure of having to do well in all these different games in order to have a high survival probability. An organism that has a successful strategy in almost all areas save one can still be fatally flawed, as the overall fitness of the organism gets contributions from the payoffs of all these different simultaneous games.

For instance, an organism must play a "game" against potential predators, and another against potential territorial rivals, and yet another against potential mates. However, this is just the tip of the iceberg. Inside the organism its immune system is also playing many simultaneous games, with white blood cell and antibody "players" competing against invading pathogens for example. Some of these games are carried out in series and some in parallel. Our immune system implements a strategy to ward off dangerous micro-organisms. It can do this while at the same time, the brain and cardio-vascular system are functioning trying to ward off a potential predator for example. On the other hand, an individual cannot simultaneously fight off a predator and look for a mate. So, in some problems a "specialist" approach is taken – the heart pumps blood, the lungs transfer oxygen, the stomach digests food etc. – while in others a "generalist" approach is taken. The latter is the rule when there is no sub-system that can specialize, such as in the case of seeking a mate and warding off predators. A human can choose to do one or the other, but not both at the same time.

The specialist approach allows for multi-tasking, a property that clearly lends a tremendous evolutionary advantage. Imagine if your immune system worked only by explicit conscious effort while you weren't distracted doing something else! Evolution is very much a phenomenon driven by the development of abilities to multi-task. In the context of multi-tasking there can be many combinations of strategies that lead to more or less the same fitness with compensations between one element and another. The multi-tasking inherent in biological systems is hierarchical in nature, with the goal of evolutionary survival being associated with an overall strategy that itself is composed of a number of building block strategies, which in their turn are composed of other less functionally complex strategies. In a dynamic where novel building block strategies can be combined at a higher level then it is possible to have an emergence of truly novel strategies that were not originally present.

## 9. Functional and Structural Modularity and Building Blocks

Above I emphasized that the overall evolutionary strategy of biological organisms can be thought of as being composed of a very large number of "building block" strategies that work together to give an overall one. These building block strategies however, are, in general, carried out in parallel not in series. This requires modularity, a manifestation of the philosophy of divide-and-conquer. Modularity allows for parallel strategy implementation and for specialized strategies. Thus, lungs take care of the fact that oxygen is necessary for cellular metabolism, converting fuel molecules, such as glucose, into biochemical energy, while the stomach and intestines take care of digestion of food, breaking it down into more easily useable molecular forms. To locate food sources, sensory organs are used, among other things, to process environmental stimuli, the brain and nervous system interpreting it and the muscles then being used to move to a food source. Each of these "modules" is highly specialized to a particular task. The cell itself also has its corresponding specialized modules all working in parallel – nucleus, mitochondria, centriole, vesicle, lysosole etc.

Thus, in this context, a module is really just a building block where we can readily its function. It shouldn't be surprising that there should be such a close relationship between building blocks as discrete structures and building blocks of strategies, a structural building block in complex systems being associated with a particular function. In fact the modules are really just the effective degrees of freedom of the system. A characteristic of a complex system is that there exists a hierarchy of modules at different scales where modules at larger scales are composed of more microscopic building block modules. Usually, we think of effective degrees of freedom as being characterized as collective excitations, being composed of combinations of the microscopic degrees of freedom. Here, I am saying that this hierarchy can and should also be thought of in terms of what the effective degrees of freedom "do", i.e., the strategy that they represent. A strategy at a higher level is then an emergent phenomenon that results from a combination of strategies at a lower level. Imagine a system and two strategies – A and B – that both enhance the fitness of the system. The question then is: Is it better to spend some time doing A and some time doing B? Or to develop the capacity to carry out both simultaneously by modularizing and using part of the system to do one and part the other? The evidence from biology is that there is no uniform answer, as examples of both occur, but that it is clearly the case that with base level functions, such as respiration, metabolism, excretion, reproduction, temperature control, immune response etc. the modular route is the only feasible one.

## 10. Can we measure complexity?

I have argued that, although manifestly complex systems are associated with a rich, emergent hierarchy of building blocks at different scales, this fact, in and of itself, is not sufficient to identify them as being complex. Instead, I argued that complexity was manifest in the special way these building blocks interacted and that this in its turn was a consequence of the system being required to fulfill a certain goal – give meaning in the context of language (symbolic complexity), and fitness in the context of biological systems (physical complexity).

If complexity is to be a meaningful scientific concept, however, then it should be measurable, or at least related to measurable quantities. What measuring device could detect hierarchical interactions between building blocks across multiple length scales? One possible measurement apparatus that springs to mind is illustrated in Figure 5 and is in itself highly complex – the human brain.

Returning to the subject of language: Imagine that we take the sentence: "To be or not to be that is the question" and we use different brains to "measure" it. Of course, there will be a significant difference in the response between an English speaker and a non-English speaker that surely would be manifest in the internal configurations of the measuring apparatus – i.e., the neural states of the two different brains when presented with the sentences. However, it is not difficult to imagine that there would also be a difference in the responses between a native English speaker and a non-native English speaker, or differences due to cultural or socio-demographic factors.

Why would there be differences? Due to two sources: non-semantic structural differences and, more importantly, semantic differences. The former is associated with the fact that an English speaker would recognize a coherent structure in the sentence irrespective of its semantic content. Beyond that, the differences in how different measuring apparatus react are more subtle, being associated with the present state of the apparatus, which, in its turn, depends on its past history.

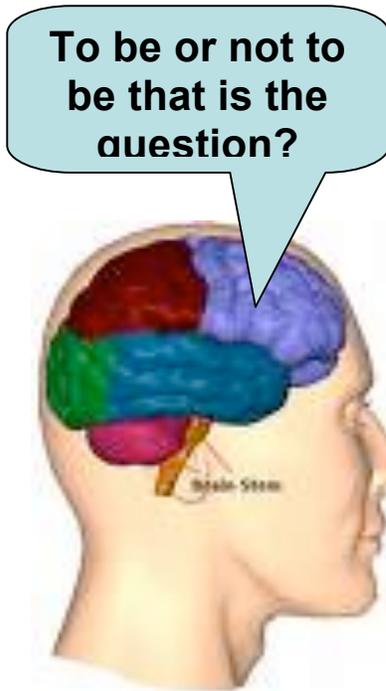

Figure 5: Is this the only apparatus capable of measuring complexity?

Someone with an intimate knowledge of Hamlet and the works of Shakespeare, and the historical and cultural context of the play, would have a different neural response to someone who was unknowledgable on all those counts.

So, does such "subjectivity" make a nonsense of using the brain as a measuring apparatus for language? Of course not! Measurement is to do with examining correlations between a property of a system and a property of a measuring device. One chooses a measuring device carefully for a given phenomenon so that this correlation is clear and readily interpretable and reproducible. In this context the brain of a non-English speaker is not that useful for measuring the semantic content of a text in English. Similarly, the brain of a non-Japanese speaker is not so useful for measuring the semantic content of a Japanese text. However, in both contexts the measuring apparatus, the brain, will respond. What differs is the nature of that response. The difference is that the individual measuring apparatus have been calibrated to respond to different signals – one to English and the other to Japanese. Is that so different to what happens in physics though?

In Figure 6 we illustrate this by considering the case of two physical phenomena: boiling water and the electrical impulses of the brain as manifest in an electro-encephalogram. The two corresponding measuring apparatus are a thermometer and a voltmeter. The thermometer is an apparatus that has been calibrated to respond to temperature and the voltmeter to potential differences. However, if we throw the voltmeter in the boiling water it will certainly respond! Just not in a way that has been calibrated and gives us the information we want or expect about the underlying system.

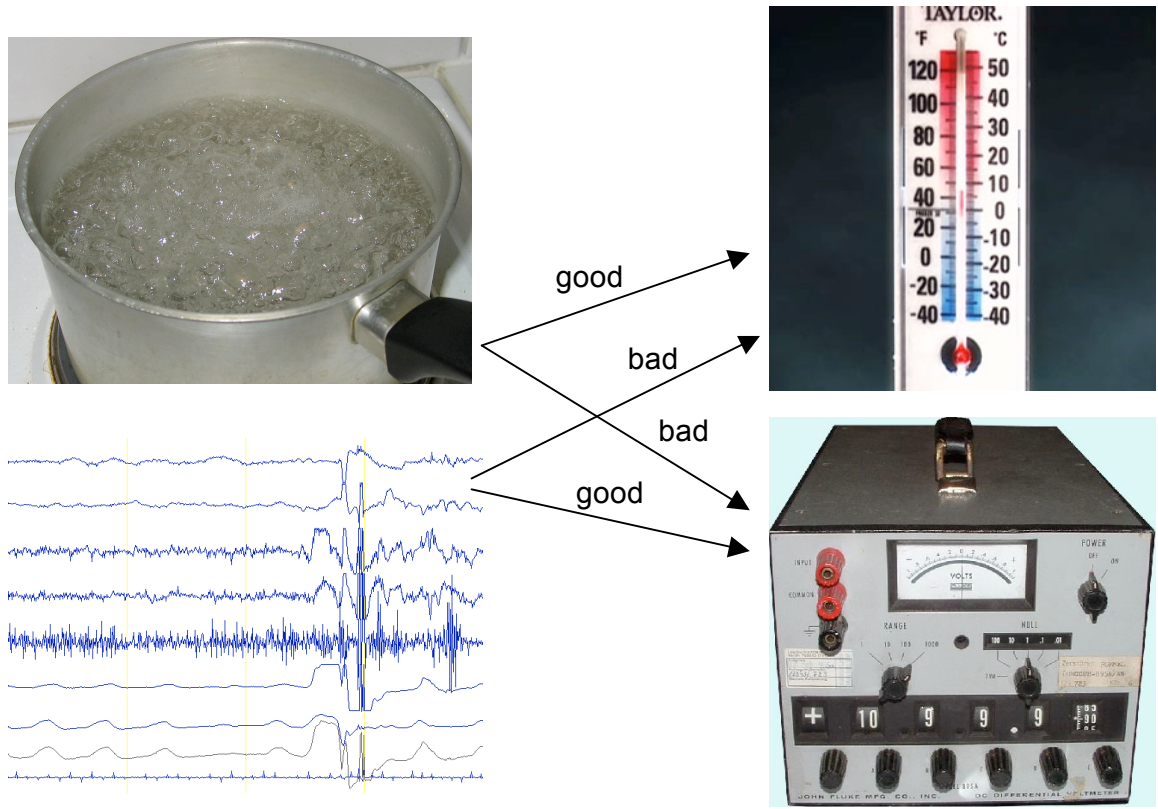

Figure 6: Correlations between different physical phenomena and some potential measuring apparatus.

The key property of the brain in terms of measuring structure and meaning in text is that it does so at multiple scales while simultaneously integrating information from those different scales. It does this by utilizing memory. Thus, the brain detects patterns irrespective of semantic meaning as, for instance, the sentient scientist sensitive to several significant sources of signal will notice in the alliterative tone. This ability to measure and integrate data across multiple scales using memory is what allows the brain to cope with the complexity inherent in the semantic content of human language and therefore measure meaning.

It is important to distinguish between the hierarchical building block structure intrinsic to language that is independent of semantics, as measured, for example, in terms of correlation functions etc., and the extrinsic complexity associated with its semantic meaning, which is an implicit relation between the system, text say, and a measuring apparatus, a brain. However, as emphasized, different brains have different calibrations.

Of course, brains are plastic and can be recalibrated, in order to measure meaning in other languages. Is there anything else that can measure meaning besides the human brain? That meaning requires a complex measuring apparatus such as the brain is manifest in the difficulties of automatic translation. Below we see different translations of the phrase "To be or not to be that is the question." as generated by an on-line translation engine. In order to test whether the system is generating any sense of "meaning" one would hope that its output from English to language X when fed in as input for translation from language X to English should return the same phrase, or a very close approximant.

1) To be or not to be that is the question.

2) Para ser o no ser que es la pregunta.

3) Om te zijn of te zijn niet dat de vraag is.

4) あるためまたはないため質問である

5) Because of a certain or because it is not, it is question?

6) Because or it is not for the sake of, that having asked and being convinced?

7) Being not to be for the sake of, or that that, you ask, are convinced?

8) It is that without having for the sake of, or, you ask, are convinced?

2) and 3), along with 1), form a closed cycle for Spanish and Dutch respectively. This happens because in this case a "literal" translation is possible, as the structures of the languages are sufficiently similar that a purely grammatical map in a single fairly simple sentence is sufficient to generate something that approximates the same semantic content. Rest assured this would not be the case if the full soliloquy was fed into the system. 4) is the translation of 1) into Japanese and 5) the output from feeding 4) into the Japanese to English direction. 6), 7) and 8) are further iterations of this process. Obviously, an unstable state has been reached!

# 11. Can we model complexity and is it the same as modeling complex systems?

Let's consider a simple mathematical model [14]: a group of point particles with position vectors $c_i$ and velocities $v_i$ and a force due to interactions with other particles which has a direction $\hat{d}_i = d_i / |d_i|$ for the $i$th particle, where

$$d_i = -\sum_{j \neq i} \frac{c_j(t) - c_i(t)}{|c_j(t) - c_i(t)|} + \sum_{j=1}^{n} \frac{v_j(t)}{|v_j(t)|} \qquad (1)$$

The first term on the right hand side represents a repulsion between the particles, while the second term an attraction that tries to align their motion. The force is made stochastic by adding a small random number to it. The effect of the force is to align the direction of a particle with $d_i$.

We can ask whether a simple mathematical model for point particles based on the interactions of (1) exhibits any of the characteristics that we have argued discriminate between complex and non-complex systems. There is certainly no manifest appearance of any of the symptoms, such as a hierarchy of effective degrees of freedom, or the emergence of strategy as a more meaningful description of an element of the system. On the contrary, the model seems to describe a very simple system with competing short-range repulsion and longer-range attraction familiar in physics. However, equation (1) has been successfully used to model the dynamics of fish shoals! Now, no one would think that fish shoals do not represent a complex system, so how can this simple system model a complex system when we have argued that the model shows none of the most discriminating features of complexity? Maybe the criteria are too restrictive? What if we use the model to describe interactions between point particles rather than fish? Does the model still represent a complex system? If so, then there are many simple physical systems that we should now classify as complex.

The resolution to this conundrum is the following: *describing complexity is not the same as describing a particular facet of a complex system*. The facet the above model describes is the *mechanistic* dynamics of how pelagic fish shoal, which just so happens to be the same type of model as describes how particles with repulsive and attractive forces interact. But fish are not mechanistic in the same sense as particles in a force field. Fish do many interesting things besides shoal. The above model cannot describe any of these other characteristics. Of course, it wasn't meant to. The model provides a mechanistic description of a behavior, a strategy – fish shoaling. We can hypothesize, given the nature of the system, that the fish do this for a reason. That it has some evolutionary advantage, such as helping reduce the predation rate. Do molecules that can be described by a similar model employ a strategy too? Of course not. One could imagine other fish behaviors that might also yield to such a mechanistic description based on a simple model. Imagine that we could, in fact, do that for any fish behaviour. Would that mean that the fish were describable in terms of a set of simple models, one per behavior? Once again, of course not. Such models do not and cannot give us any insight into complexity or the notion of what is "complex" in a complex system. They can however, give us quantitative models of certain aspects of complex systems that, importantly, can, in principle, be compared with experiment.

So complex systems can be modelled, in the above sense, but that does not mean we can model complexity. Is this possible? Well, we certainly do not have any existing theory that can do the job. What ingredients would such a model need? Well, for physical complexity, i.e. biological systems, first, it would have to be a model that worked at the level of strategies and states not just at the level of a unique dynamical law evolving a state. Second, it would have to function at a level where the strategies were not a priori known, nor was their payoff. This also has to be an emergent property. The systems that most closely approach this paradigm are agent-based systems, such as are used to model financial markets [le baron]. In such a case, one models strategies in a context where there is no explicit pre-specified payoff function. Rather, the success of an agent strategy is implicitly dependent on the strategies of all other agents, and therefore cannot be calculated until at a given moment of time all the other strategies have been specified. To understand these systems, except in the case of the very simplest strategies, such as random trading, inevitably, a simulation has to be run. Although such systems exhibit several features that I would claim are important to the development of complexity they also leave several important ones out. For instance, in such systems there is no building block hierarchy of strategies. A reason for this is that there is no real need for modularity, i.e. no associated multi-tasking.

## 12. From symbolic to physical complexity

I have used throughout as paradigms of complexity human languages and biological organisms. In section 2. I compared a text of human language to some texts of "genetic" language, posing the question of whether the latter was also complex. It seems fairly clear that it is genetic language that leads to physical complexity through development. How then can such physical complexity be reached if the language that underlies it is not itself symbolically complex? As argued in the case of human language, this implies that it is complex from the demands of genetic "semantics" not genetic "grammar". In other words strands of DNA, on their own, in solution, say, are not complex, just as Japanese is not complex for me, as I cannot measure the complex interactions inherent in it due to the demands that it have meaning. My brain has not been calibrated for Japanese. As stated, meaning comes from an interaction between a text and a brain understood as a measuring apparatus. What is the "brain" for genetic texts? DNA in solution has no meaning relative to its surroundings – the solvent. However, DNA in the context of a cell does have meaning, as is manifest by the complex activity of the cell. In this case the cell environment acts as the equivalent of the brain, a measuring apparatus giving meaning to what is written in the genetic texts. Once this meaning is established, then an action can be taken, such as to produce a particular protein with an eventual macro-characteristic at the level of a phenotypic trait. This is precisely the genotype-phenotype map. Unfortunately, our understanding of it is minimal.

So, in the cell there is an instruction booklet which needs to be read and, more importantly, needs to be understood. This instruction booklet exhibits a high degree of complexity in the context of a cell machinery that is capable of understanding it. These instructions are subsequently acted upon and this eventually leads to some phenotypic traits, which in their turn can be understood as facilitating certain strategies. Thus, in the case of birds, there are instructions for making wings and feathers and these allow the bird to implement the evolutionarily useful strategy of flying.

## 13. Conclusions

In this contribution I have given a very personal account of what I believe complexity, and the related notion of complex system, to be. Starting off with the premise that biological systems and human languages are definitely complex I tried to determine what non-tautological properties distinguished those systems from others. The idea was not to provide a rigorous definition of these concepts but rather see what phenomenological properties discriminated most. Properties such as many degrees of freedom, non-linear interactions and Edge-of-Chaos definitely do not discriminate. Neither does the more sophisticated concept of a hierarchy of effective degrees of freedom – building blocks – as a function of scale, as simple physical systems also exhibit such hierarchies.

What does seem to distinguish the complex systems I consider is the property that there exist emergent characteristics – meaning and fitness – that induce interactions across different levels of building block and are such as to require an integration of the contributions from the different building blocks across different levels. This is quite distinct to physical systems where building block structure on one scale is effectively frozen out at others. It is for that reason that the physical sciences have been so successful when compared to the biological ones. In the physical world pretty much things are either homogeneous, crystals etc, or random, e.g. glass or a gas. It is the stability of the physical world that leads to a relative lack of diversity in structure and function. It takes a lot of energy to break up nuclei, atoms, molecules etc. For complexity we need to be able to construct a hierarchy of building blocks that is neither too stable nor too unstable. We need low energy to do that.

We also need to avoid the tyranny of physical law. Complex systems were argued to be characterized more by what they do rather than what they are. Physical systems, on the other hand, are completely constrained, in a usually quite transparent way, as to what they can do. Complex systems however, are best described in terms of different strategies that they can implement. Put metaphorically - complex systems are verbs while physical systems are nouns. The tyranny of physical law is strongly related to the simple requirements that such systems have to obey – find the state of least energy, or least action, for example. In contrast, complex systems are characterizable by a building block hierarchy of function wherein physical law acts only as a constraint to the possible dynamics not as an explanation of the dynamics per se as it does in physical systems.

The escape from the tyranny of physical law has come about due to the challenge of evolving in a complex environment. This has driven systems to develop structural building block hierarchies that, in their turn, represent functional building block hierarchies. Each functional building block can then be tasked with solving a part of the puzzle of survival.

The conclusion then is that one can isolate properties of biological systems and human languages that do seem to truly distinguish them when compared to any physical system. In that sense we can take these properties as characteristic of complexity and complex systems. Whether there are systems, other than the ones mentioned here remains to be seen. Of course, the reader may disagree with my usage of the word complexity as it is far more restrictive than previously used meanings of the word. It is, after all, just a word. What is more important than the word, is the set of properties that I have put under the rubric of this word, and our ability to measure and model them.